\documentclass[12pt]{article}
\usepackage{oldgerm}
\usepackage{yfonts}
\usepackage{euler}
\usepackage{graphics}
\usepackage{bm}
\usepackage{graphicx}
\usepackage{epstopdf}
\usepackage{amsmath}
\usepackage{amssymb}
\usepackage{amstext}
\usepackage{amscd}
\usepackage{amsfonts}
\usepackage{appendix}
\usepackage{color}
\usepackage{wasysym}

\newcommand{\be}{\begin{equation}}
\newcommand{\ee}{\end{equation}}
\newcommand{\ben}{\begin{eqnarray}}
\newcommand{\een}{\end{eqnarray}}

\newcommand{\nd}{\noindent}

\title{{\bf Strong correlations between the exponent $\alpha$
and the particle number  for a Renyi-monoatomic gas in
Gibbs' statistical mechanics}}
\author{A. Plastino, M. C. Rocca\\
La Plata National University\\
and Argentina's National Research Council,\\
(IFLP-CCT-CONICET)-C. C. 727, 1900\\
La Plata - Argentina}

\date{\today}

\begin{document}

\maketitle

\begin{abstract}

Appealing to the 1902 Gibbs' formalism for classical statistical mechanics (SM), the first SM axiomatic theory ever that successfully 
explained equilibrium thermodynamics, we will here show that 
already at the classical level there is a strong correlation
between the Renyi's exponent $\alpha$ and the number of
particles for very simple systems. No reference to heat
baths is needed for such a purpose.
\vskip 3mm 
\nd Keywords: Gibbs theory, Entropy, Renyi's monoatomic gas;  Harmonic oscillators\\
\nd PACS: {05.20.-y, 05.70.Ce, 05.90.+m}

\end{abstract}

\renewcommand{\theequation}{\arabic{section}.\arabic{equation}}

\section{Introduction}

In his celebrated book of 2002, ELEMENTARY PRINCIPLES OF STATISTICAL MECHANICS \cite{gibbs}, Gibbs put forward an axiomatic 
theory for statistical mechanics (SM) (the first one indeed of that kind) that was able to microscopically and successfully explain equilibrium thermodynamics. He invented the ensemble notion. All this happened 60 years {\it before the advent of MaxEnt}. One can certainly work on classical SM without appeal to MaxEnt, and this is what we are going to do here. Why? Because MaxEnt workings with q-generalized entropies have recently received serious questioning \cite{q1,q2}, and we wish to disentangle our findings from MaxEnt.

Renyi's  information measure $S_R$ generalizes  both
 Hartley's and Shannon's  ones, quantifying our ignorance regarding a system's
structural features. $S_R$ is considered an important
quantifier in variegated areas of science, for instance,  ecology, quantum information,  
Heisenberg's XY spin chain model, theoretical computing,
conformal field theory, quantum quenching, diffusion processes,
etc. \cite{0,1,2,3,4,5,6,7,8,9,10,11}. Also, the Renyi entropy is important in  statistics as indicator of diversity.    Here we tackle, using the classical statistical  mechanics of Gibbs, the simplest conceivable systems, the ideal gas, and an ensemble of independent harmonic oscillators,
 showing  that a strange phenomenology emerges even in these trivial scenarios. In particular a strong correlation between
 Renyi's exponent $\alpha$ and the particle number 
 emerges, without appeal to the heat bath notion.

\subsection{Gibbs postulates}

They were advanced in 1902 and constituted the first ever axiomatics for classical 
statistical mechanics (SM) \cite{gibbs,libro}. 
   Gibbs refers to a phase space location as the "phase" $z$ of the system \cite{libro}. He also introduced the notion of ensemble. The following
statements completely explain in microscopic fashion the corpus of classical
equilibrium thermodynamics \cite{libro}.\vskip 3mm \nd
1) The probability that at time $t$ the system will be found in the dynamical
state characterized by $z$ equals the probability $P(z)$ that a system
randomly selected from the ensemble shall possess the phase $z$.
2) All phase-space neighborhoods (cells) have the same a priori probability.
3) The ensemble's probability  $P(q_1,\ldots,q_n,p_1,\ldots,p_n,t)$ (self-explanatory notation) depends only upon the system's Hamiltonian $H(q_1,\ldots,q_n,p_1,\ldots,p_n,t)$ in an exponential (negative) fashion.
4) The time-average of a dynamical quantity $F$ equals its average over the
ensemble, evaluated using   $P$.

\vskip 3mm \nd The prevalent contemporary SM-axiomatics, suitable for quantum mechanics, is that of Jaynes 
\cite{jaynes}, usually called MaxEnt, {\it which is not employed here}.  Now, it is well known that,  for entropies like Renyi's, the second Gibbs' postulate has to be amended by replacing the exponential function by the so called q-exponential one $e_q$ (for $e_q$'s properties, see \cite{t2})

\be e_\alpha(-x)= [1- (1-\alpha)x]^{1/(1-\alpha)},\,\,\, e_\alpha(-x) \rightarrow e^{-x}\,\,\,
for\,\,\,\alpha \rightarrow 1, \ee
and then, in the canonical ensemble, 

\be    \label{formita} P(H) \sim e_\alpha(-\beta H), \,\,\,\beta=1/T, \ee with $T$ the temperature.

\subsection{Renyi's measure}

Renyi's $S_R$ is defined as \cite{0}:
\begin{equation}
\label{eq2.14} S_R=\frac {1} {1-\alpha}\ln\left( \int\limits_M
P^{\alpha}d\mu\right),
\end{equation}
and the accompanying (canonical ensemble) Gibbs'  probability
distribution $P$ is given by \ref{formita}. The general form of the partition function was derived by Gibbs in 1902 [see   \cite{gibbs}, Eq. (92)]. If $Z$ stands for Renyi's
partition function, one has 

\begin{equation}
\label{eq2.13}
Z=\int\limits_M
\left[1+(1-\alpha)\beta H\right]^{\frac {1}
{\alpha-1}}d\mu,
\end{equation}
\begin{equation}
\label{eq2.12} P=\frac {1} {Z}
\left[1+(1-\alpha)\beta H\right]^{\frac {1} {\alpha-1}}.
\end{equation}
Herefrom we denote the classical energy
 by $U$ and its mean value by $<U>$, and insist upon the fact that MaxEnt is NOT appealed to.

\subsection{Renyi's and Tsallis' measures}

\nd It is well known \cite{t2} that $S_R$ is intimately linked to Tsallis' entropy $S_{\alpha}$ 
\cite{t2,t1,t3}. In fact, one easily ascertains that, given

\be \label{talfa} {\cal T}(\alpha) = \int P^{\alpha} d\mu, \ee

\be \label{tsallisent} S_{\alpha} = \frac{1}{1-\alpha}[{\cal T}(\alpha) - 1] d\mu,\ee
then

\be \label{SRSQ}    S_R(\alpha)  = \frac{1}{1-\alpha} \ln{[(1-\alpha)S_{\alpha}+1]}.\ee

\nd Note that the  Tsallis' canonical probabilities are also q-exponentials \ref{eq2.12} \cite{t2,t1}. Accordingly, Tsallis results, within Gibbs' tenets,   will necessarily coincide with those of  Renyi's. 
 Anything to be found below for the later will automatically be also valid in the Tsallis scenario as well.

\setcounter{equation}{0}

\section{Renyi's partition function for the monoatomic ideal gas}

\nd Using appropriate units, the partition function of
$\nu-$dimensional monoatomic gas of $n$ particles is ($[0 < \alpha \le 1]$), after an adequate Gibbs' treatment 
is  

\begin{equation}
\label{eq2.1}
Z=V^n\int\limits_{-\infty}^{\infty}
\left[1+\frac {\beta(1-\alpha)} {2m}(p_1^2+p_2^2+\cdot\cdot\cdot
p_n^2)\right]^{\frac {1}{\alpha-1}}d^{\nu}p_1d^{\nu}p_2\cdot\cdot
\cdot d^{\nu}p_n,
\end{equation}
with $V$ the volume. Using spherical coordinates in a space of  
 $\nu n$ dimensions, the above  integral becomes
\begin{equation}
\label{eq2.2} 
Z=\frac {2\pi^{\frac {\nu n} {2}}V^n}
{\Gamma\left(\frac {\nu n} {2}\right)}
\int\limits_{0}^{\infty}
\left[1+\frac {\beta(1-\alpha)} {2m} p^2
\right]^{\frac {1}{\alpha-1}}p^{\nu n-1}dp.
\end{equation}
We have integrated over the angles and taken
$p^2=p_1^2+p_2^2+\cdot\cdot\cdot p_n^2$. Changing variables in the fashion  $x=p^2$, the
last  integral is
\begin{equation}
\label{eq2.3}
Z=\frac {\pi^{\frac {\nu n} {2}}V^n}
{\Gamma\left(\frac {\nu n} {2}\right)}
\int\limits_{0}^{\infty}
\left[1+\frac {\beta(1-\alpha)} {2m} x
\right]^{\frac {1}{\alpha-1}}x^{\frac {\nu n-2} {2}}dx,
\end{equation}
so that, appealing to ref.\cite{ts1}, we are led to 
\begin{equation}
\label{eq2.4} 
Z=V^n\left[\frac {2\pi m} {\beta(1-\alpha)}\right]^{
\frac {\nu n} {2}}
\frac {\Gamma\left(\frac {1} {1-\alpha}-\frac {\nu n} {2}\right)}
{\Gamma\left(\frac {1} {1-\alpha}\right)}.
\end{equation}
Note that the $\Gamma$'s argument in the numerator 
can not vanish. Thus, $\frac {1} {1-\alpha}-\frac {\nu n} {2}>0$
(strictly).

In similar fashion one evaluates the mean energy. One has 
\[Z<U>=V^n\int\limits_{-\infty}^{\infty}
\left[1+\frac {\beta(1-\alpha)} {2m}(p_1^2+p_2^2+\cdot\cdot\cdot
p_n^2)\right]^{\frac {1}{\alpha-1}}\]
\begin{equation}
\label{eq2.5}
\frac {p_1^2+p_2^2+\cdot\cdot\cdot p_n^2} {2m}
d^{\nu}p_1d^{\nu}p_2\cdot\cdot
\cdot d^{\nu}p_n.
\end{equation}
Integrating over the angles we find

\begin{equation}
\label{eq2.6} 
Z<U>=\frac {\pi^{\frac {\nu n} {2}}V^n}
{m\Gamma\left(\frac {\nu n} {2}\right)}
\int\limits_0^{\infty}
\left[1+\frac {\beta(1-\alpha)} {2m} p^2
\right]^{\frac {1}{\alpha-1}}p^{\nu n+1}dp.
\end{equation}
Now, using  $x=p^2$ once more we are led to

\begin{equation}
\label{eq2.7}
Z<U>=\frac {\pi^{\frac {\nu n} {2}}V^n}
{2m\Gamma\left(\frac {\nu n} {2}\right)}
\int\limits_0^{\infty}
\left[1+\frac {\beta(1-\alpha)} {2m} x
\right]^{\frac {1}{\alpha-1}}x^{\frac {\nu n} {2}}dx.
\end{equation}
 At this point, we use again  ref.\cite{ts1} and get 
\begin{equation}
\label{eq2.8} 
<U>=\frac {V^n} {Z}\frac {\nu n} {2\beta(1-\alpha)}
\left[\frac {2\pi m} {\beta(1-\alpha)}\right]^{\frac {\nu n} {2}}
\frac {\Gamma\left(\frac {1} {1-\alpha}-\frac {\nu n} {2}
-1\right)} {\Gamma\left(\frac {1} {1-\alpha}\right)}.
\end{equation}
One replaces now in this last result the value encountered above for  $Z$   and obtain
\begin{equation}
\label{eq2.9} 
<U>=\frac {\nu n} {\beta[2\alpha-\nu n(1-\alpha)]}.
\end{equation}
Finally, the derivative with respect to  $T$ yields for the specific heat 
at constant volume

\begin{equation}
\label{eq2.10} 
C_V=\frac {\nu nk} {2\alpha-\nu n(1-\alpha)}.
\end{equation}

\setcounter{equation}{0}

\section{Limits to the particle-number for  Renyi's monoatomic gas}

The original content of the present communication emerges from an analysis of the Gamma functions 
involved in evaluating $Z$ and $<U>$. Remember that $[0 <\alpha \le 1]$. According to (2.4), 
the integral  (\ref{eq2.1}) converges and becomes both  positive and finite for  
\begin{equation}
\label{eq3.1} 
\frac {1} {1-\alpha}-\frac {\nu n} {2}>0.
\end{equation}
Analogously, according to (2.8), we need 
\begin{equation}
\label{eq3.2} 
\frac {1} {1-\alpha}-\frac {\nu n} {2}-1>0.
\end{equation}
These two conditions immediately set severe limitations on the particle-number $n$ that read
\begin{equation}
\label{eq3.3} 
1\leq n<\frac {2\alpha} {\nu(1-\alpha)}.
\end{equation}
There is a maximum permissible number of particles.
For instance, if 
$\alpha=1-10^{-2}, \nu=3$, we have
\begin{equation}
\label{eq3.4} 
1\leq n<66.
\end{equation}
No more than 65 particles are allowed. Keeping the dimensionality 3, for $\alpha=1/2$ only ONE particle is allowed! Even worse, for $\alpha=1/4$, NO particles exist. Roughly, to have a number of particles of the order of $10^n$,
 one needs $\alpha$ of the order of $1-10^{-n}$.
Note that (\ref{eq3.3}) implies $\alpha>\frac {\nu n} {2+\nu n}$.
See Fig 1.
\newpage
\begin{figure}[h]
\begin{center}
\includegraphics[scale=0.6,angle=0]{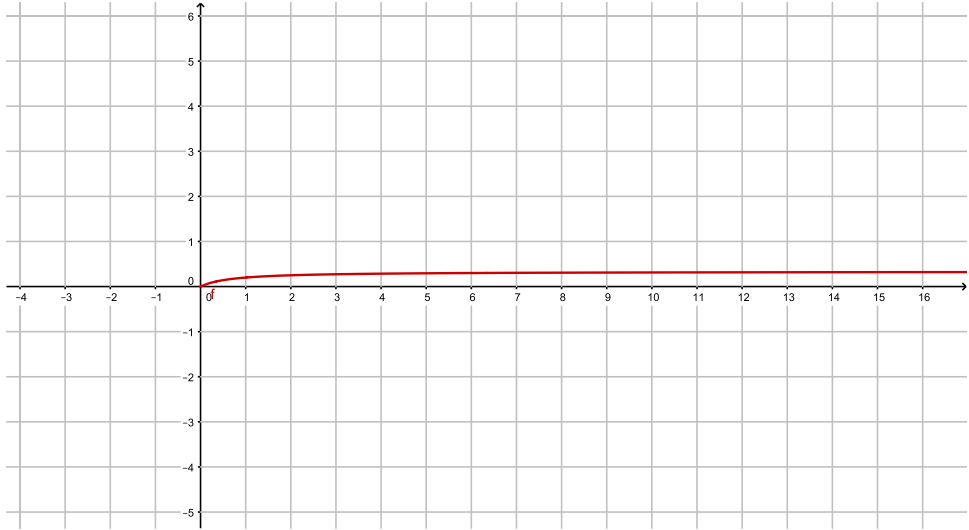}
\vspace{-0.2cm} \caption{In the 3-dimensional
case, we plot $\frac {n} {2+3n}$ vs $n$.
The permitted values for $\alpha$ must lie above the
represented curve.}\label{fig1}
\end{center}
\end{figure}

\setcounter{equation}{0}

\section{Renyi's partition function for $n$ independent  harmonic oscillators}

\nd Using appropriate units, the partition function 
for n $n-$set of independent $\nu-$dimensional Harmonic Oscillators is
given by
\[Z=\int\limits_{-\infty}^{\infty}
\left[1+\beta(1-\alpha)(p_1^2+p_2^2+\cdot\cdot\cdot p_n^2+
q_1^2+q_2^2+\cdot\cdot\cdot q_n^2)\right]^{\frac {1}{\alpha-1}}\]
\begin{equation}
\label{eq4.1}
d^{\nu}p_1d^{\nu}p_2\cdot\cdot\cdot d^{\nu}p_n
d^{\nu}q_1d^{\nu}q_2\cdot\cdot\cdot d^{\nu}q_n.
\end{equation}
We appeal again to spherical coordinates and integrate over the angles. One has
\begin{equation}
\label{eq4.2} 
Z=\frac {2\pi^{\nu n}}
{\Gamma\left(\nu n\right)}
\int\limits_0^{\infty}
\left[1+\beta(1-\alpha) p^2
\right]^{\frac {1}{\alpha-1}}p^{2\nu n-1}dp,
\end{equation}
where
$p^2=p_1^2+p_2^2+\cdot\cdot\cdot p_n^2+
q_1^2+q_2^2+\cdot\cdot\cdot q_n^2$.
We repeat the variables' change  $x=p^2$ and are led to 

\begin{equation}
\label{eq4.3}
Z=\frac {\pi^{\nu n}}
{\Gamma\left(\nu n\right)}
\int\limits_0^{\infty}
\left[1+\beta(1-\alpha) x
\right]^{\frac {1}{\alpha-1}}x^{\nu n-1}dx.
\end{equation}
From  ref.\cite{ts1} we gather that 
\begin{equation}
\label{eq4.4} 
Z=\left[\frac {\pi} {\beta(1-\alpha)}\right]^{\nu n}
\frac {\Gamma\left(\frac {1} {1-\alpha}-\nu n\right)}
{\Gamma\left(\frac {1} {1-\alpha}\right)},
\end{equation}
and
\[Z<U>=\int\limits_{-\infty}^{\infty}
\left[1+\beta(1-\alpha)(p_1^2+p_2^2+\cdot\cdot\cdot p_n^2+
q_1^2+q_2^2+\cdot\cdot\cdot q_n^2)\right]^{\frac {1}{\alpha-1}}\]
\begin{equation}
\label{eq4.5}
(p_1^2+p_2^2+\cdot\cdot\cdot q_n^2)
d^{\nu}p_1d^{\nu}p_2\cdot\cdot\cdot d^{\nu}p_n
d^{\nu}q_1d^{\nu}q_2\cdot\cdot\cdot d^{\nu}q_n.
\end{equation}
With spherical coordinates this becomes 

\begin{equation}
\label{eq4.6} 
Z<U>=\frac {2\pi^{\nu n}}
{\Gamma\left(\nu n\right)}
\int\limits_0^{\infty}
\left[1+\beta(1-\alpha) p^2
\right]^{\frac {1}{\alpha-1}}p^{2\nu n+1}dp, 
\end{equation}
and after setting $x=p^2$,
\begin{equation}
\label{eq4.7}
Z<U>=\frac {\pi^{\nu n}}
{\Gamma\left(\nu n\right)}
\int\limits_0^{\infty}
\left[1+\beta(1-\alpha) x
\right]^{\frac {1}{\alpha-1}}x^{\nu n}dx.
\end{equation}
Recourse again to ref.\cite{ts1} yields
\begin{equation}
\label{eq4.8} 
<U>=\frac {1} {Z}\frac {\nu n} {\beta(1-\alpha)}
\left[\frac {\pi} {\beta(1-\alpha)}\right]^{\nu n}
\frac {\Gamma\left(\frac {1} {1-\alpha}-\nu n
-1\right)} {\Gamma\left(\frac {1} {1-\alpha}\right)}.
\end{equation}
Replacing $Z$ above we reach, finally,
\begin{equation}
\label{eq4.9} 
<U>=\frac {\nu n} {\beta[\alpha-\nu n(1-\alpha)]},
\end{equation}
and

\begin{equation}
\label{eq4.10} 
C=\frac {\nu nk} {\alpha-\nu n(1-\alpha)}.
\end{equation}

\setcounter{equation}{0}

\section{Limits to the number of independent harmonic oscillators}

\setcounter{equation}{0}

Finiteness of  $Z$ entails
\begin{equation}
\label{eq5.1} 
\frac {1} {1-\alpha}-\nu n>0,
\end{equation}
and for  $<U>$ 
\begin{equation}
\label{eq5.2} 
\frac {1} {1-\alpha}-\nu n-1>0.
\end{equation}
Thus, a new maximum for $n$ ensues 
\begin{equation}
\label{eq5.3} 
1\leq n<\frac {\alpha} {\nu(1-\alpha)},
\end{equation}
and for 
$\alpha=1-10^{-2}, \nu=3$

\begin{equation}
\label{eq5.4} 
1\leq n<33.
\end{equation}
We see also that $1>\alpha> 2/3$ for having a single-HO system.  
$1>\alpha> 5/6$   is the condition for having a system of two oscillators, 
$1>\alpha> 8/9$ the condition for having a system of three HO's, etc.

\setcounter{equation}{0}

\section{Discussion}

We have been working within Gibbs' classical
scheme for statistical mechanics. No appeal to MaxEnt was made. 
It was seen that demanding  finiteness of the partition function 
and mean energy severely 
limits the number $n$ of independent components of a system in a 
Renyi scenario (and also in
a Tsallis nonadditive one). There are  bounds for $n$. 
The most bizarre situation is encountered for some $\alpha$-values that do not 
permit the system's existence because $n$ can nor exceed zero.

\nd These problem imply that there exists a strong correlation 
between $\alpha$ and the number of particles. 
This fact has been proposed in Refs. \cite{varios},-for instance.
Although  Renyi always considered $\alpha$ to be 
an independent parameter, one might argue that our present results do give additional impetus to the $\alpha  -  n$-correlation  proposal, and thus deserve dissemination. Fig. 1 clearly illustrates
this $\alpha-n$ correlation. $\alpha$ must lie above the curve
drawn there.

\section*{Acknowledgments}

\nd The authors acknowledge support from CONICET (Argentine
Agency).

\end{document}